\documentclass[twocolumn,preprint,3p]{elsarticle}

\usepackage{multicol}
\usepackage{color}
\usepackage{graphicx}
\usepackage{graphicx,epstopdf}
\usepackage{dcolumn}
\usepackage{bm}
\usepackage{mathrsfs}
\usepackage{soul}
\usepackage{mathtools}
\usepackage{blkarray, bigstrut}
\usepackage{hyperref}
\usepackage{color,soul}
\usepackage{mathtools}

\usepackage{graphics}
\usepackage{amssymb}
\usepackage{amsmath}
\usepackage{mathtools}
\usepackage{color}
\journal{Journal of Theoretical Biology}

\usepackage{graphicx}
\usepackage{epsfig}
\usepackage{multirow}
\usepackage{epstopdf}
\usepackage{subfigure}
\usepackage{mathtools}
\usepackage{hyperref}
\usepackage{graphicx}
\usepackage[space]{grffile}
\usepackage{latexsym}
\usepackage{textcomp}
\usepackage{longtable}
\usepackage{tabulary}
\usepackage{booktabs,array,multirow}
\usepackage{amsfonts,amsmath,amssymb}
\usepackage{epstopdf}
\usepackage{comment}
\usepackage[table]{xcolor}
\usepackage{notoccite}
\arrayrulecolor{blue}

\begin{document}
	
\begin{frontmatter}	
	
	\title{Eco-evolutionary cyclic dominance among predators, prey, and parasites}

	\author[a]{Sayantan Nag Chowdhury}
	\author[b]{Jeet Banerjee}
	\author[c,d,e,f,g]{Matja{\v z} Perc}	
	\author[h]{Dibakar Ghosh\corref{cor1}}
	\ead{diba.ghosh@gmail.com}
	\cortext[cor1]{Corresponding author}

	\address[a]{Department of Environmental Science and Policy, University of California, Davis, CA 95616, USA}
	\address[b]{BYJU’S, Think \& Learn Pvt. Ltd., IBC Knowledge Park, 4/1 Bannerghatta Main Road, Bangalore 560029, India}
    \address[c]{Faculty of Natural Sciences and Mathematics, University of
		Maribor, Koroška cesta 160, 2000 Maribor, Slovenia}
	\address[d]{Alma Mater Europaea, Slovenska ulica, 17, 2000 Maribor, Slovenia}
    \address[e]{Department of Medical Research, China Medical University
		Hospital, China Medical University, Taichung, Taiwan}
	\address[f]{Complexity Science Hub Vienna, Josefstädterstraße 39,
		1080 Vienna, Austria}
	\address[g]{ Department of Physics, Kyung Hee University, 26 Kyungheedae-ro,
	Dongdaemun-gu, Seoul, Republic of Korea}
	\address[h]{Physics and Applied Mathematics Unit, Indian Statistical Institute,
		203 B. T. Road, Kolkata 700108, India}

\begin{abstract}	
Predator-prey interactions are one of ecology's central research themes, but with many interdisciplinary implications across the social and natural sciences. Here we consider an often-overlooked species in these interactions, namely parasites. We first show that a simple predator-prey-parasite model, inspired by the classical Lotka--Volterra equations, fails to produce a stable coexistence of all three species, thus failing to provide a biologically realistic outcome. To improve this, we introduce free space as a relevant eco-evolutionary component in a new mathematical model that uses a game-theoretical payoff matrix to describe a more realistic setup. We then show that the consideration of free space stabilizes the dynamics by means of cyclic dominance that emerges between the three species. We determine the parameter regions of coexistence as well as the types of bifurcations leading to it by means of analytical derivations as well as by means of numerical simulations. We conclude that the consideration of free space as a finite resource reveals the limits of biodiversity in predator-prey-parasite interactions, and it may also help us in the determination of factors that promote a healthy biota.
\end{abstract}
\end{frontmatter}
	
\section{Introduction}	
 Predator-prey interaction in theoretical ecology brings several applicable mathematical models to the limelight, leading to a possible treasure trove of information. Thomas Robert Malthus first proposed a significant development in this direction by incorporating exponential growth in a single species model \cite{malthus2007essay}. Despite its various shortcomings, this simple model provides a fertile building block for predator-prey interactions and triggers further fundamental discoveries. By refining this model, a plethora of systems like the classical Lotka–Volterra predator-prey model \cite{lotka1925elements,volterra1926variazioni} and, later, the Rosenzweig–MacArthur model \cite{rosenzweig1963graphical}, including density-dependent prey growth and a functional response, are formed. Most of these models are formed by incorporating different realistic essence in the system, and thus, those systems are capable of offering diverse emergent dynamics. However, the contribution of free space toward the predator-prey competitive relationship is relatively ignored in the existing literature. 
\par Free space provides every species an opportunity to thrive; however, it never anticipates any benefit for helping others. Any individual can use the free space for their well-being. Recently researchers brought this altruistic behavior of the free space to the limelight by investigating its impact on the evolution of cooperation \cite{chowdhury2021complex,chowdhury2021eco,roy2022eco}. Various simpler models with diverse motivations \cite{helbing2009outbreak,nag2020cooperation,jiang2010role,chowdhury2020distance,meloni2009effects,Sar_2022,noh2004random,aktipis2004know,vainstein2007does,chowdhury2019synchronization,smaldino2012movement} have been proposed to study the impact of free space on natural and human-made systems. Nevertheless, how free space's unselfish concern to benefit others than itself influences the predator-prey interaction is yet to be discovered. To investigate this, we initially resort to a mathematical model where predators depend on a particular organism, prey for living. A predator feed prey and preys feed the insect parasites. These insect parasites consume food only from predators. We formulate a set of differential equations by considering this simple cyclical interaction. Unfortunately, this simplified model can not stabilize the parasites; hence an unrealistic scenario occurs. In the absence of any physically realistic result, we begin investigating the interaction among the prey, predator, and parasite using the game's theoretical tools. Our constructed model based on the payoff (interaction) matrix offers various notable dynamics in the form of steady-state and periodic oscillation. The cyclic interaction among the species allows cyclical dominating one another under favorable circumstances. Examples of cyclic dominance \cite{szolnoki2014cyclic} in nature are already well-documented. 
\par The spontaneous emergence of cyclic dominance is found in several ecological setups involving microbial populations \cite{nahum2011evolution,kerr2002local}, plant systems \cite{lankau2007mutual,durrett1998spatial}, and marine benthic systems \cite{jackson1975alleopathy}. There are ample real-life examples like the genetic regulation in the repressilator \cite{elowitz2000synthetic}, the mating strategy of side-blotched lizards \cite{sinervo1996rock}, oscillating frequency of lemmings in a simple vertebrate predator–prey community \cite{gilg2003cyclic} and the oscillating frequency of the Pacific salmon \cite{guill2011three} highlight the beauty of cyclical interactions to maintain the sustainable biodiversity in nature. Interactions among living organisms are much more complicated compared to the interactions among particles; hence, it is essential to understand how cyclically competing strategies promote natural biodiversity. {For the study of cyclical interactions, the classical rock-paper-scissors game} \cite{hofbauer1998evolutionary,nowak2006evolutionary} {has proven to be an effective tool. This evolutionary game entailing cyclic dominance with a few simple microscopic rules can capture the essence of several realistic, complex spatial patterns} \cite{reichenbach2007noise,he2010spatial,kabir2021role,reichenbach2007mobility}. In the present article, we derive a simple set of ordinary differential equations based on the ecological interactions between predator, prey, and parasites. Since the nonlinear model formulated using the fundamental principles offers biologically unrealistic and mathematically unstable dynamics, our approach of inclusion the selfless contribution of free space not only brings the evolutionary cycling as a likely outcome of the eco-evolutionary model but also can capture a more realistic description of the competitive ecological models.
\par The section-wise organization of this article is as follows. In Sec.\ \eqref{Model}, we investigate a three-dimensional dynamical system based on the cyclical interactions among predator, prey, and parasite motivated by the Lotka-Volterra model, which is one of the central paradigms for the emergence of periodic oscillations in nonlinear systems. Unfortunately, these nonlinear equations fail to capture any realistic description, as parasites are unable to stabilize in our constructed model (See Supplementary material Sections (1-4)). Therefore, we take the help of the game's theoretical tools and are able to devise an eco-evolutionary model offering a more realistic description of predator-prey-parasite interactions. We aim to shed light on the impact of altruistic behavior of the free space, and hence we consider the generous contribution of free space in the payoff (interaction) matrix. We elaborately outline the model's main properties (existence, uniqueness, positivity). Motivated by Refs.\ \cite{hauert2006evolutionary,gokhale2016eco}, we assume each subpopulation dies at a respective constant death rate and explore the system's dynamics numerically with the variation of these parameters in Sec.\ \eqref{Results}. We provide sufficient numerical evidence to validate the emergence of cyclic dominance. Finally, we briefly summarize our findings in Sec.\ \eqref{Conclusion} and round off by providing an outlook on the challenges and promising future research efforts.
\section{Mathematical model} \label{Model}	
 We consider a simplistic scenario where predators consume preys and preys eat up some insect parasites for their survival. The insect parasite consumes food from the predator's body at the expense of the predators. Thus, we have a cyclical interaction between predator, prey, and parasite. To formulate this cyclical interaction, initially, we start with the following system of ordinary differential equations, 		
		\begin{equation} \label{1}
			\begin{split}
				\dfrac{dx}{dt^{'}}=rx-d_1x-d_2xy+e_1d_3xz,\\
				\dfrac{dy}{dt^{'}}=e_2d_2xy-d_4y-d_5yz,\hspace{0.8cm} \\ 
				\dfrac{dz}{dt^{'}}= e_3d_5yz-d_6z-d_3xz. \hspace{0.8cm}
			\end{split}
		\end{equation}
Here, the prey, predator, and parasites' biomass are given by $x$, $y$, and $z$, respectively. We assume that the prey population grows linearly with the intrinsic growth rate $r$ without predators and parasites. $d_1$ is the natural death rate of prey, while $d_2$ is the rate of predation of prey by the predator. $e_1$ is a dimensionless quantity representing the conservation efficiency for converting parasites' biomass to prey's biomass. We consider the predator's response as a Holling type I functional response. Here, $e_2$ is the conversion efficiency (dimensionless) for converting prey to predator's biomass. $d_4$ is the natural death rate of predators, and $d_5$ is the death rate of the predator due to parasitism. $e_3$ is the conversation efficiency (dimensionless) for converting predator's biomass to insect parasite's biomass, and $d_6$ is the natural death rate of insect parasite. $t^{'}$ is used here to represent the time. Here, the dimension of $r,d_1,d_4$, and $d_6$ is $\dfrac{1}{\text{time}}$, and that of $d_2$ and $d_5$ is $\dfrac{1}{\text{time} \times \text{biomass}}$.
\par We use the following set of transformations to introduce a new set of nondimensionalized parameters
\begin{equation} \label{2}
			\begin{split}
				u=\dfrac{d_2}{r}x,
				v=\dfrac{d_2}{r}y,
				w=\dfrac{e_1 d_3}{r}z,
				t=rt^{'},
				\alpha=\dfrac{d_1}{r},\\
				\beta=\dfrac{d_4}{e_2 r},
				\gamma=\dfrac{d_5}{e_1e_2d_3},
				\epsilon=e_2,\\
				\delta=\dfrac{e_3d_5}{d_2},
				\xi=\dfrac{d_6}{r},\hspace{0.2cm}\text{and}\hspace{0.2cm}
				\eta=\dfrac{d_3}{d_2}.
			\end{split}
		\end{equation}		
Using these parameters, we get the nondimensionalized system as follows		
		\begin{equation} \label{3}
			\begin{split}
				\dot{u}=\dfrac{du}{dt}=u(1-\alpha-v+w),\\
				\dot{v}=\dfrac{dv}{dt}=\epsilon v(u-\beta-\gamma w),\hspace{0.25cm} \\ 
				\dot{w}=\dfrac{dw}{dt}=w(\delta v-\xi-\eta u).\hspace{0.2cm}
			\end{split}
		\end{equation}
Here, $\alpha$ and $\beta$ are the nondimensionalized intrinsic death rates of prey and predator, respectively. $\epsilon$ is the time-scale separation between the life-span of prey and predator populations, which belongs to $(0,1]$. $\gamma$ is the death rate of predators due to parasitism, and $\delta$ is the growth rate of insect parasites due to parasitism. $\xi$ is the natural death rate of the insect parasite, and $\eta$ is the death rate of insect parasite due to consumer by prey. Clearly, all these parameters of the model \eqref{3} are positive for the physically meaningful interpretation of the system. A detailed analysis of this system is given in the Supplementary Sections (1-4), and we show that the parasites cannot stabilize if we consider this system. Hence, we resort to the game theoretical approach to obtain a biologically implementable model.		
\par To construct the payoff matrix from the system \eqref{3}, we observe the following points
		\begin{itemize}
			\item Interaction coefficient between predator-prey, incurred by prey (coefficient of $uv$ in $\dot{u}$) is $-1$.
			\item Interaction coefficient between prey-parasite, incurred by prey (coefficient of $uw$ in $\dot{u}$) is $+1$.
			\item Interaction coefficient between predator-prey, incurred by predator (coefficient of $uv$ in $\dot{v}$) is $\epsilon$.
			\item Interaction coefficient between predator-parasite, incurred by predator (coefficient of $vw$ in $\dot{v}$) is $-\epsilon\gamma$.
			\item Interaction coefficient between predator-parasite, incurred by parasite (coefficient of $vw$ in $\dot{w}$) is $\delta$.
			\item Interaction coefficient between prey-parasite, incurred by parasite (coefficient of $uw$ in $\dot{w}$) is $-\eta$.	
		\end{itemize}
	Note that we do not consider any intraspecific interactions as the model \eqref{3} does not contain any terms like $u^2$, $v^2$, and $w^2$. Hence, the payoff matrix in the absence of any intraspecific competition looks like
	\[
	\begin{blockarray}{cccc}
		\hspace{0.1 cm} & \text{Prey}  & \text{Predator} & \text{Parasite} \\
		\begin{block}{c(ccc)}
			\text{Prey} \hspace{0.1 cm} & 0 & -1 & 1  \\
			\text{Predator} \hspace{0.1 cm} & \epsilon & 0 & -\epsilon\gamma \\
			\text{Parasite} \hspace{0.1 cm}  & -\eta & \delta & 0 \\
		\end{block}
	\end{blockarray}
	\]
	Now, we consider the contribution of free space to all other populations. Let $\psi_1$, $\psi_2$, and $\psi_3$ be the reproductive benefits towards prey, predator and parasite, respectively. Despite giving such selfless promotion of others' welfare, free space does not receive any positive benefits. Thus, the modified payoff matrix after including free space as the fourth interacting entity can be described as
		\[
	\begin{blockarray}{ccccc}
		\hspace{0.1 cm} & {p}  & {q} & {r} & {s} \\
		\begin{block}{c(cccc)}
			{p} \hspace{0.1 cm} & 0 & -1 & 1 & \psi_1 \\
			{q} \hspace{0.1 cm} & \epsilon & 0 & -\epsilon\gamma & \psi_2\\
			{r} \hspace{0.1 cm}  & -\eta & \delta & 0 & \psi_3 \\
			{s} \hspace{0.1 cm}  & 0 & 0 & 0 & 0\\
		\end{block}
	\end{blockarray}
	\]
	Here, $p$, $q$, $r$ and $s$ denote the fraction of prey, predator, parasite, and free-space, respectively. Clearly, the total population surrounded by the free-space is one, i.e., 
	\begin{equation} \label{13}
		\begin{split}
			p+q+r+s=1.
		\end{split}
	\end{equation}
     Using the payoff matrix, we can derive the fitness of prey, predator, parasite, and free space as follows
     \begin{equation} \label{14}
     	\begin{split}
     		f_p=-q+r+\psi_1 s, \hspace{2.8cm}\\
     		f_q=\epsilon p -\epsilon \gamma r +\psi_2 s, \hspace{2.6cm}\\
     		f_r=-\eta p +\delta q +\psi_3 s,\hspace{2.5cm}\\
     		f_s=0. \hspace{4.7cm}
     	\end{split}
     \end{equation}
 			\begin{figure*}[!t]
 	\centerline{\includegraphics[width=1.10\textwidth]{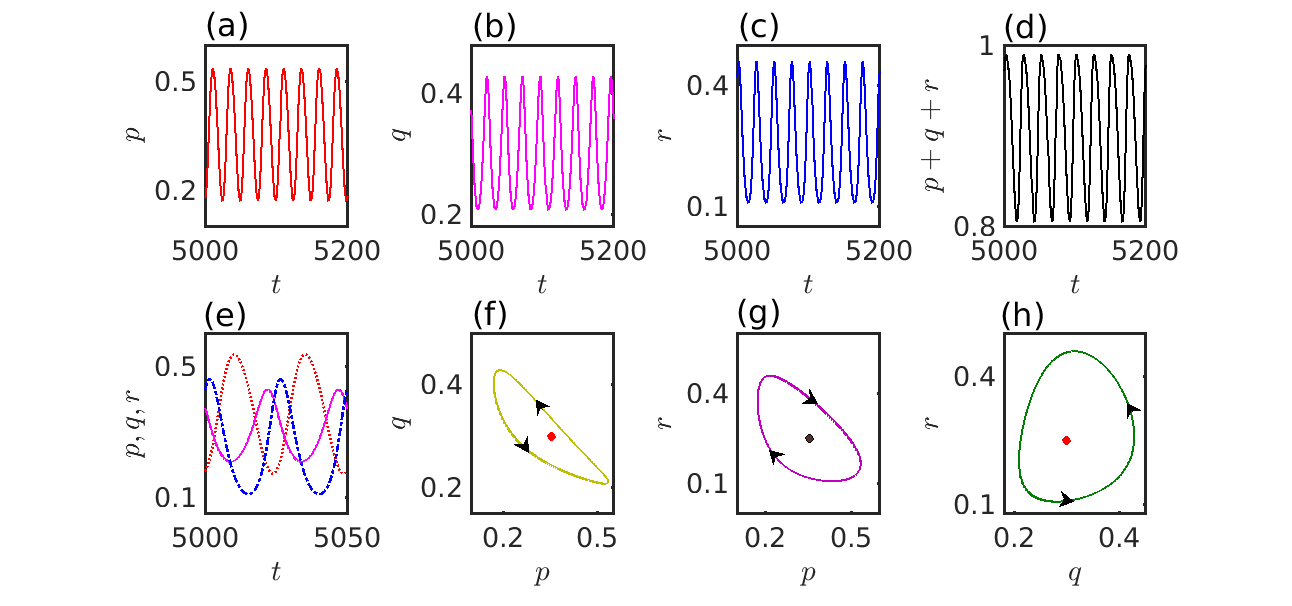}}
 	\caption{{\bf Cyclic dominance of prey, predator and parasite}: While the system \eqref{3} does not facilitate the coexistence of all species, the eco-evolutionary model \eqref{16} allows the survivability of all three simultaneously. The time evolutions of (a) prey $p(t)$, (b) predator $q(t)$ and parasite (c) $r(t)$  maintain periodic dynamics, and their overall population density (d) $p+q+r$ lies within $[0,1]$. This boundedness makes the dynamics biologically interpretable. (e) The periodic dynamics provide each species an opportunity over a suitable time window for maintaining dominance over the other two. The projections of the trajectory on the two-dimensional phase spaces (f) $p-q$, (g) $p-r$, and (h) $q-r$ are presented the cyclic dominance. We further plot the interior equilibrium $(0.3552,0.2993,0.2492)$ with a filled circular marker. It is a saddle-focus for this particular parameter set. The arrows describe the motion of a particle along the closed orbit. Parameters: $\alpha=1.0$, $\beta=0.8$, $\epsilon=0.1$, $\gamma=1.0$, $\delta=1.39$, $\xi=0.42$, $\eta=0.1$, $\psi_1=0.52$, $\psi_2=0.72$, and $\psi_3=0.41$. Initial condition: $(0.3,0.3,0.3)$.}
 	\label{Fig2}
 \end{figure*}
	    By observing the coefficients of $u$, $v$, and $w$ ignoring the signs from the system \eqref{3}, the natural death rates of prey, predator, and parasite are $1-\alpha$, $\epsilon\beta$, and $\xi$, respectively. Thus, we have the eco-evolutionary model as follows,
	    \begin{equation} \label{15}
	    	\begin{split}
	    		\dot{p}=p(f_p-(1-\alpha)),\\
	    		\dot{q}=q(f_q-\epsilon\beta),\hspace{0.8cm}\\
	    		\dot{r}=r(f_r-\xi). \hspace{0.9cm}
	    	\end{split}
	    \end{equation}
    
    Using Eq.\ \eqref{13}, the simplified model looks like

    \begin{equation} \label{16}
    	\begin{split}
    		\dot{p}=p(-\psi_1 p - (1+\psi_1)q+(1-\psi_1)r+(\psi_1+\alpha-1)),\hspace{0.5cm}\\
    		\dot{q}=q((\epsilon-\psi_2)p-\psi_2 q-(\epsilon\gamma+\psi_2)r+(\psi_2-\epsilon\beta)),\hspace{1.1cm}\\
    		\dot{r}=r(-(\eta+\psi_3)p+(\delta-\psi_3)q-\psi_3r+(\psi_3-\xi)),\hspace{1.2cm}\\
    		\dot{s}=-\dot{p}-\dot{q}-\dot{r}. \hspace{6.3cm}
    	\end{split}
    \end{equation}

This model has ten parameters $\alpha$, $\beta$, $\gamma$, $\eta$, $\delta$, $\xi >0$, $\psi_1$, $\psi_2$, $\psi_3 \geq 0$, and $\epsilon \in (0,1]$. To ensure that the constructed model is biologically well-behaved, we investigate the positivity of the model for initial densities $(p_0,q_0,r_0)$, where $p_0$, $q_0$, $r_0 \geq 0$. Since the right-hand side of Eq.\ \eqref{16} is a polynomial, it is continuous and locally Lipschitz. Thus, the solution of this proposed system \eqref{16} with initial conditions $(p_0\geq0,q_0\geq0,r_0\geq0)$ must exist and is unique in the interval $[0,\infty) \times [0,\infty) \times [0,\infty)$. Note that the overall initial densities must satisfy the constraint $0 \leq p_0+q_0+r_0 \leq 1$ for a biologically meaningful interpretation. Furthermore from the eco-evolutionary model \eqref{16} with non-negative initial conditions $(p_0\geq0,q_0\geq0,r_0\geq0)$, we have

	\begin{figure*}[!t]
	\centerline{\includegraphics[width=1.00\textwidth]{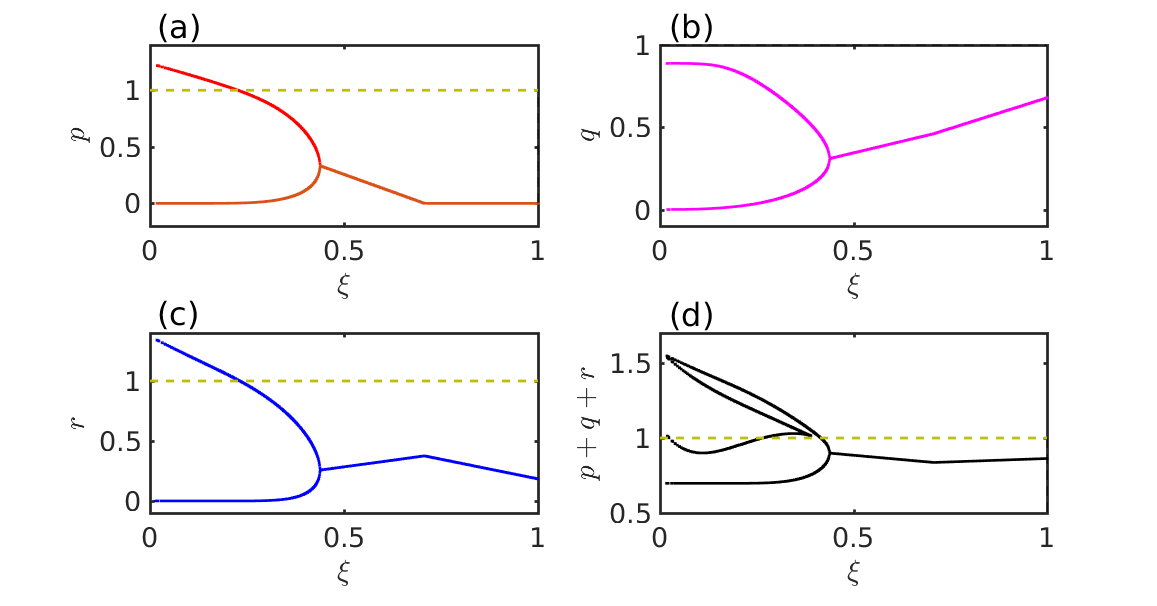}}
	\caption{{\bf Influence of parasite's natural death rate}: The bifurcation diagrams by varying the parasite's natural death rate $\xi$ for (a) prey $p(t)$, (b) predator $q(t)$ and parasite (c) $r(t)$ are shown. Subfigure (d) portrays that the overall population $p+q+r$ lies within the bounded interval $[0,1]$ for $\xi \in [0.4160,1]$. The vertical dashed line indicates $p+q+r=1$. The system \eqref{16} experiences a Hopf bifurcation at $\xi=0.43793199$. Beyond this value of $\xi$, the system converges to a steady state. The prey population (red) goes extinct as $\xi$ increases, while the increment of $\xi$ benefits the predators (magenta). The parasites (blue) suffer due to the increased natural death rate $\xi$. Consequently, predators are able to get extra aid, and hence, their density increases. This enhancement is also further supported by the free space induced benefits $\psi_2>\psi_1$ and $\psi_2>\psi_3$. We use the FORTRAN-90 compiler and iterate the system \eqref{16} using the RKF45 method with $700000$ iterations, out of which $690000$ iterations are discarded as transient. The integrating step length is fixed $h=0.01$. Parameters: $\alpha=1.0$, $\beta=0.8$, $\epsilon=0.1$, $\gamma=1.0$, $\delta=1.39$, $\eta=0.1$, $\psi_1=0.52$, $\psi_2=0.72$, and $\psi_3=0.41$. Initial condition: $(0.3,0.3,0.3)$. }
	\label{Fig3}
\end{figure*}

    \begin{equation} \label{16-new}
	\begin{split}
		p(t)=p_0\exp\bigg[\int_{0}^{t}(-\psi_1 p(s) - (1+\psi_1)q(s)\\+(1-\psi_1)r(s)+(\psi_1+\alpha-1))ds\bigg]\geq0,\\
		q(t)=q_0\exp\bigg[\int_{0}^{t}((\epsilon-\psi_2)p(s)-\psi_2 q(s)\\-(\epsilon\gamma+\psi_2)r(s)+(\psi_2-\epsilon\beta))ds\bigg]\geq0,\\
		r(t)=r_0\exp\bigg[\int_{0}^{t}(-(\eta+\psi_3)p(s)+(\delta-\psi_3)q(s)\\-\psi_3r(s)+(\psi_3-\xi))ds\bigg]\geq0, \forall t\geq0.
	\end{split}
\end{equation}	

This confirms $p(t)$, $q(t)$, $r(t) \geq 0$ for all $t\geq 0$. Hence, for each non-negative initial density $(p_0,q_0,r_0)$ with $0 \leq p_0+q_0+r_0 \leq 1$, the proposed deterministic model \eqref{16} has a unique positive solution $(p(t),q(t),r(t))$ for all $t \geq 0$.

\begin{figure*}[!t]
	\centerline{\includegraphics[width=1.10\textwidth]{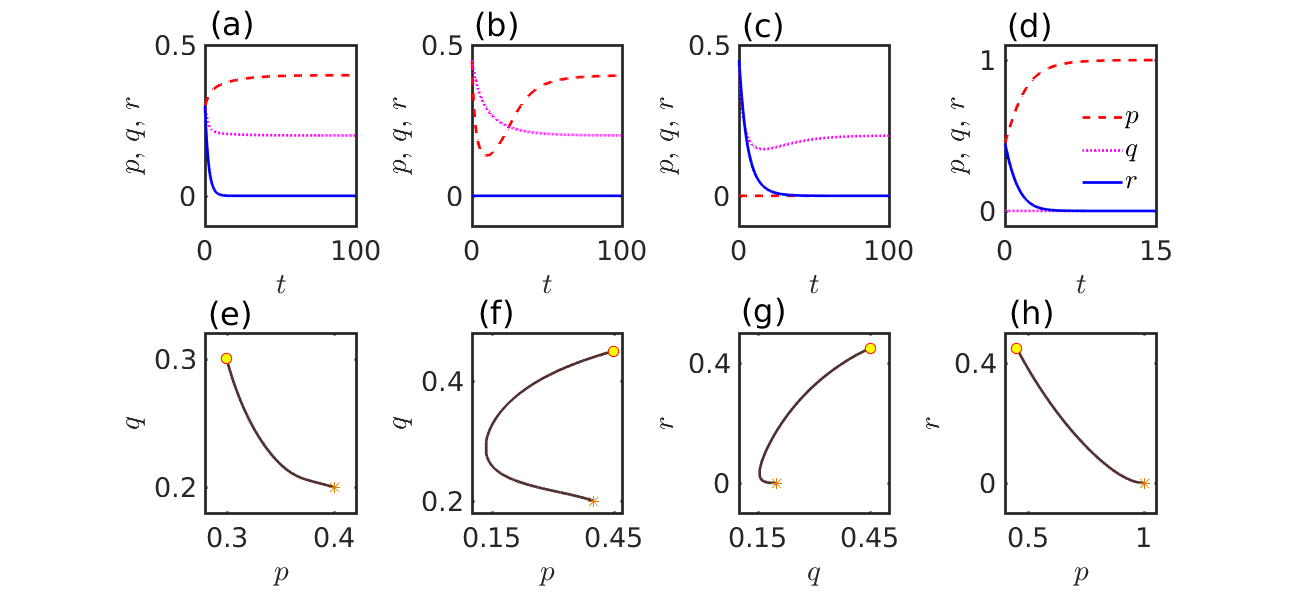}}
	\caption{{\bf Multiple steady states for different initial conditions}: We choose four different initial densities (a, e) $(0.3,0.3,0.3)$, (b, f) $(0.45,0.45,0)$, (c, g) $(0,0.45,0.45)$, and (d, h) $(0.45,0,0.45)$.  The parasite will die in the long run for all these initial conditions. Each column represents the same result. The top row depicts the temporal evolution, and the bottom row portrays the two-dimensional phase space projection of the trajectories. The first two columns show the coexistence of the prey and predator. The third column delineates the predator's sole survivability, as the prey's initial density is zero. The fourth column reflects that the prey can only survive in the asymptotic state. The yellow circular marker stands for the two-dimensional projection of the chosen initial conditions. The star marker describes the two-dimensional projection of the converged stationary point. All parameters are kept fixed at $\alpha=1.0$, $\beta=0.8$, $\epsilon=0.3$, $\gamma=0.4$, $\delta=0.7$, $\xi=0.5$, $\eta=0.6$, $\psi_1=0.5$, $\psi_2=\psi_3=0.3$.}
	\label{Fig4}
\end{figure*}			
\section{Results} \label{Results}
 The main drawback of the model \eqref{3} is that the parasites cannot stabilize in society, even if they exist. We consider the model \eqref{16} to overcome this issue. Initially, we fix all the parameters' values at $\alpha=1.0$, $\beta=0.8$, $\epsilon=0.1$, $\gamma=1.0$, $\delta=1.39$, $\xi=0.42$, $\eta=0.1$, $\psi_1=0.52$, $\psi_2=0.72$, and $\psi_3=0.41$. Although the system possesses eight distinct stationary points; however, only five stationary points exist for the parameter set mentioned above. The extinction equilibrium $(0,0,0)$ is a saddle as the eigenvalues of the Jacobian of the system \eqref{16} at origin are $\lambda_1=-0.01$, $\lambda_2=0.52$ and $\lambda_3=0.64$. The predator-parasite free stationary point is $(1,0,0)$. The eigenvalues of the Jacobian at this point are $\lambda_1=-0.52$, $\lambda_2=0.02$ and $\lambda_3=-0.52$. Hence, it is also a saddle. The prey-parasite free stationary point $(0,0.8889,0)$ is a saddle with the Jacobian eigenvalues $\lambda_1=-0.64$, $\lambda_2=-0.8311$ and $\lambda_3=0.8611$. The prey-free stationary point is $(0,0.2463,0.5643)$ is a saddle-focus with Jacobian's eigenvalues $\lambda_{1,2}=-0.2043\pm0.3331i$ and $\lambda_3=0.4165$, where $i=\sqrt{-1}$. The interior equilibrium $(0.3552,0.2993,0.2492)$ is a saddle-focus where the eigenvalues of the Jacobian at this point are $\lambda_{1,2}=0.0045\pm0.2583i$ and $\lambda_3=-0.5114$. All other stationary points are not biologically meaningful. Thus, the system will not converge to any stationary points for the parameter set mentioned above. Under this circumstance, the system either oscillates or leaves the phase space after a finite time. We iterate the model \eqref{16} numerically using the Runge–Kutta–Fehlberg method with a fixed integration time step $h=0.01$. In fact, all the figures of this study are done using the same method and FORTRAN 90 compiler. 
\par To begin with, we choose all the species' equal densities, and without loss of generality, we select the initial condition at $(0.3,0.3,0.3)$. We plot the dynamics of the eco-evolutionary model \eqref{16} in Fig.\ \eqref{Fig2}. We find that all the species periodically dominate one another in different time windows (See Fig.\ \eqref{Fig2} (e)). While in system \eqref{3}, parasites do not even get a chance to stabilize, here all three species co-exist simultaneously in the model \eqref{16}. We notice that not only $p$, $q$, $r$ lie within $[0,1]$ (See Fig.\ \eqref{Fig2} (a-c)), but also the overall population $p+q+r$ lies too within physically implementable range $[0,1]$ (See Fig.\ \eqref{Fig2} (d)). We also plot the unstable interior equilibrium (circular marker) in Figs.\ \eqref{Fig2} (f-h). The cyclic dominance among prey, predator, and parasite is one of the exciting, realistic essence captured by our eco-evolutionary model. 
\subsection{Influence of parasite's natural death rate} For further understanding, we investigate the influence of the parameter $\xi$ in Fig.\ \eqref{Fig3}. $\xi>0$ reflects the death rate of the parasite in our proposed model \eqref{16}.
\begin{figure*}[!t]
	\centerline{\includegraphics[width=1.00\textwidth]{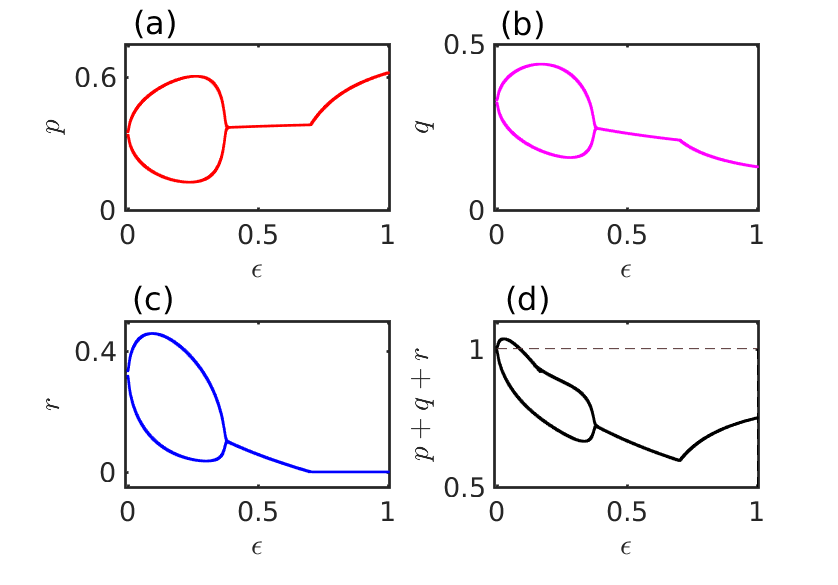}}
	\caption{{\bf The impact of $\epsilon$ on the eco-evolutionary dynamics}: Bifurcation diagram by changing $\epsilon$ for (a) $p(t)$, (b) $q(t)$ and (c) $r(t)$. Using MATCONT \cite{dhooge2003matcont}, we identify the system \eqref{16} experiencing Hopf bifurcations at two different values of $\epsilon=0.0033008651$ and $0.37323091$. The periodic solution arises at the first point, and at the second point, it disappears. Since the natural death rate of predators is a function of $\epsilon$, their density decreases with increasing $\epsilon \in (0,1]$. Consequently, parasites get less amount of food, and hence, their density diminishes. In fact, beyond $\epsilon=0.70185963$, the parasites will become extinct. This reduction of predators' density ultimately boosts the prey's density, and thus, we observe an increment in the prey's population with increasing $\epsilon$. The horizontal dashed line in subfigure (d) indicates $p+q+r=1$. Thus, the results are biologically interpretable for $\epsilon \in [0.089,1]$, so that the overall population density $p+q+r$ remains bounded within $[0,1]$. The figure is drawn by varying the $\epsilon \in (0,1]$ with step length $0.001$. Parameters: $\alpha=1.0$, $\beta=0.8$, $\gamma=1.0$, $\delta=1.39$, $\eta=0.1$, $\psi_1=0.52$, $\psi_2=0.72$, $\psi_3=0.41$, and $\xi=0.42$. Initial condition is kept fixed at $(0.3,0.3,0.3)$.}
	\label{Fig5}
\end{figure*}
We vary the parameter $\xi$ within the interval $(0,1]$ with a fixed step length $0.001$ and fixed initial condition $(0.3,0.3,0.3)$. This initial point allows all the species to have equal densities, at least in the beginning. Figure \eqref{Fig3} (d) depicts the overall population $p+q+r$ lies within [0,1] for $\xi \in [0.4160,1]$. The dashed horizontal line indicates the upper bound $p+q+r=1$ beyond which the dynamics are not meaningful from the biological point of view. Within this physically meaningful range of $\xi \in [0.4160,1]$, a period-halving bifurcation in the system. Using MATCONT \cite{dhooge2003matcont}, we have identified a local bifurcation point at $\xi=0.43793199$, where the periodic solution disappears, and the interior equilibrium stabilizes. At $\xi=0.43793199$, we find the eigenvalues of the linearized system around the interior equilibrium $(0.333,0.3094,0.2571)$ are $\lambda_{1,2}=\pm 0.2609i$ and $\lambda_3=-0.5013$. Crossing the imaginary axis in the complex plane of the pair of complex conjugate eigenvalues against the variation of $\xi$ affirms that a Hopf bifurcation occurs (See section 11.2 of Ref.\ \cite{hale2012dynamics}).
\par Figure \eqref{Fig3} is drawn with the same parameters values chosen in Fig.\ \eqref{Fig2}. We find the periodic solution and the steady states for $\xi \in [0.4160,1]$. Figure \eqref{Fig3} (a) shows the prey's density monotonically decreases and dwindles to zero with increasing $\xi$ in the steady state regime. Interestingly, the predators' density increases with increasing $\xi$ as shown in Fig.\ \eqref{Fig3} (b). This increment attests to the contribution of the other nine parameters in the complex evolutionary dynamics of our model \eqref{16}. Noticeably, free space altruistically contributes to all the species; nevertheless, $\psi_2$ is higher than $\psi_1$ and $\psi_3$ for this figure. Thus, free space favors the evolution of predators for the chosen set of parameter values. Obviously, the other parameters also play a vital part in forming the asymptotic dynamics. Figure \eqref{Fig3} (c) initially, the parasites' density increases in the steady state regime and, finally, decreases beyond a critical value of $\xi$. Note that the figure is drawn for a fixed initial condition $(0.3,0.3,0.3)$. 
\subsection{Effect of different initial conditions} The role of initial conditions in determining the final asymptotic behavior is of utmost importance. To illustrate this factor, we choose four different initial conditions $(p_0,q_0,r_0)$ maintaining the constraint $p_0+q_0+r_0=0.9$ in Fig.\ \eqref{Fig4}. Since the free-space-induced benefits are higher for the prey for our chosen parameter values, it is expected to observe the dominance of prey.  We choose four distinct initial points $(0.3,0.3,0.3)$, $(0.45,0.45,0)$, $(0,0.45,0.45)$ and $(0.45,0,0.45)$ in Fig.\ \eqref{Fig4} (a-d), respectively. In all these four subfigures, parasites die in the long run. The Fig.\ \eqref{Fig4} (a-b) show that both prey and predator survive, and in both cases, the prey's density surpasses that of the predator. Figure\ \eqref{Fig4}(c) reflects the sole survivability of the predators, while prey can only survive in Fig.\ \eqref{Fig4}(d). Since the initial densities of the parasite, prey and predator are zero in Figs.\ \eqref{Fig4}(b-d), respectively; thus there is no scope of reproduction for them in the asymptotic limit. This expectation is also demonstrated through Fig.\ \eqref{Fig4} (b-d). This Fig.\ \eqref{Fig4}  also confirms the system may converge to diverse stationary points solely depending on the initial conditions (see Supplementary Section (5)). 
\begin{figure*}[!t]
	\centerline{\includegraphics[width=1.00\textwidth]{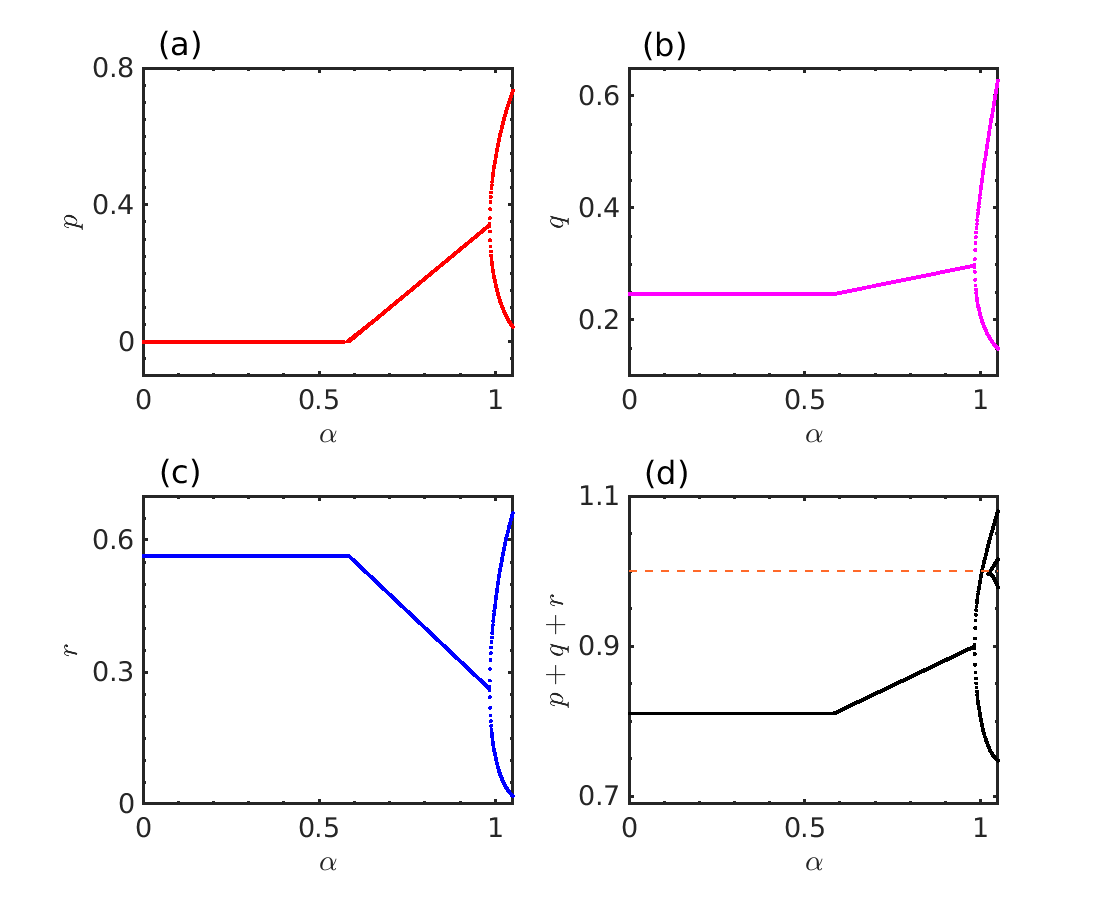}}
	\caption{{\bf Importance of $\alpha$ on the emergent dynamics}: Bifurcation diagram by changing $\alpha$ for (a) $p(t)$, (b) $q(t)$ and (c) $r(t)$. We plot here the dynamics of the eco-evolutionary model \eqref{16} by varying $\alpha \in (0,1.2]$ with fixed step length $0.001$ and fixed initial condition $(0.3,0.3,0.3)$. However, $\alpha>1$ makes the prey's natural death rate $(1-\alpha)$ negative. Thus, we are only interested in examining the dynamics for $\alpha \in (0,1]$. In fact, the overall population density $p+q+r$ lies within the bounded interval $[0,1]$ for $\alpha \in (0,1.004)$, as reflected through the subfigure (d). The horizontal dashed line indicates $p+q+r=1$. With an increasing value of $\alpha$, the death rate $(1-\alpha)$ of prey decreases. And thus, the prey (red line in subfigure (a)) can revive from their zero density at $\alpha=0.58348744$. Their density grows monotonically after this value of $\alpha$ in the steady state regime. Ultimately, this stationary point vanishes through the Hopf bifurcation at $\alpha=0.98402536$, and a periodic solution appears. As the prey increases beyond a critical value of $\alpha$, this transition affects the parasites. Their respective density (blue) thus reduces in this steady state regime as portrayed through subfigure (c). As a parasite consumes food from the predator's body, consequently, this reduction of the parasite's density will enhance the predator's density (magenta in subfigure (b)). Parameters: $\beta=0.8$, $\gamma=1.0$, $\delta=1.39$, $\eta=0.1$, $\xi=0.42$, $\epsilon=0.1$, $\psi_1=0.52$, $\psi_2=0.72$, and $\psi_3=0.41$.}
	\label{Fig6}
\end{figure*}
\subsection{Effect of time-scale separation $\epsilon$ between prey and predator populations on the eco-evolutionary dynamics} Now, we inspect the impact of $\epsilon$ on our proposed model \eqref{16}. We vary $\epsilon$ within $(0,1]$ with small step-length $0.001$ and fixed initial condition $(0.3,0.3,0.3)$ in Fig.\ \eqref{Fig5}. For fixed $\beta$, the increase of $\epsilon$'s value enhances the death rate of the predator. Thus, it is natural to observe a decreasing trend in the predators' density, particularly in the steady state regime. We find the same result in Fig.\ \eqref{Fig5} (b). Consequently, the prey's density will get the opportunity to grow in favorable circumstances. The same trend is observed in Fig.\ \eqref{Fig5} (a). In the steady state regime, the parasites decrease monotonically, and all the parasites will become extinct at $\epsilon=0.70185963$, as shown in Fig.\ \eqref{Fig5} (c). Initially, a small range $(0,0.088]$ of $\epsilon$ exists in Fig.\ \eqref{Fig5} (d), where the overall population $p+q+r$ will exceed the unity. This range is neglected for the sake of a biologically well-behaved system \eqref{16}. Using MATCONT, we identify two values of $\epsilon$ where the Hopf bifurcation arises. Out of which, the first one $\epsilon=0.0033008651$ is not subject of concern here, as for this value of $\epsilon$, we have $p+q+r>1$. At $\epsilon=0.37323091$, once again, a Hopf bifurcation occurs, the periodic solution disappears, and the interior equilibrium stabilizes. Here, the overall population $p+q+r$ lies within the domain $[0,1]$ and makes the results interpretable from the biological points of view. 
\begin{figure*}[!t]
	\centerline{\includegraphics[width=1.10\textwidth]{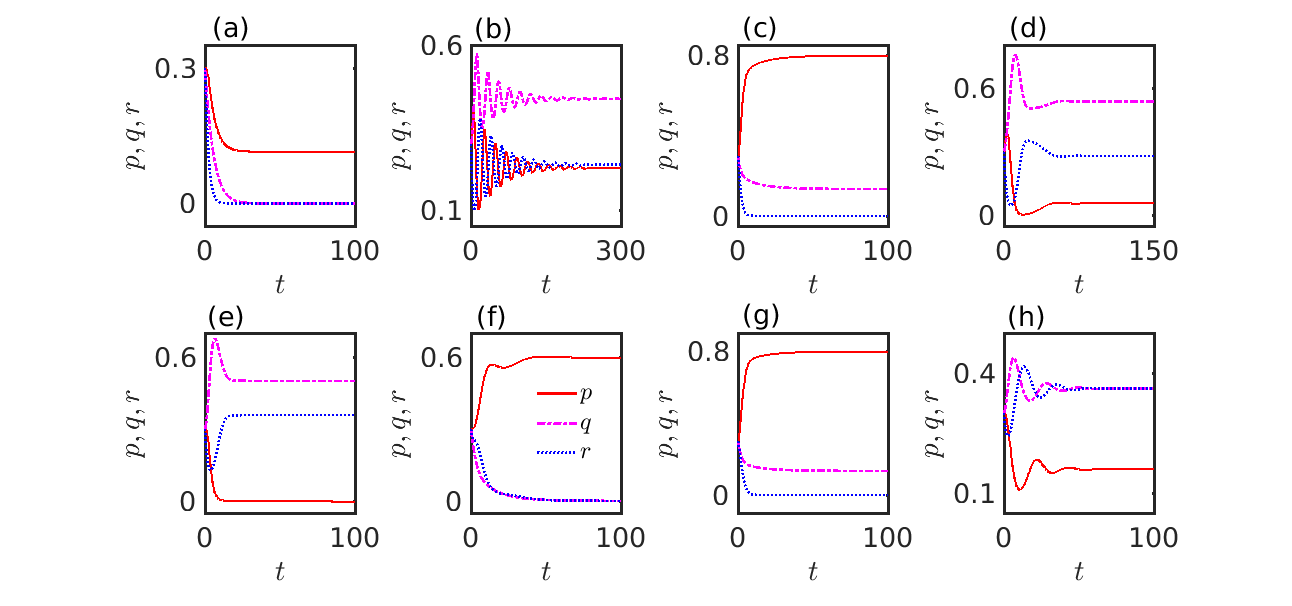}}
	\caption{{\bf Impact of altruistic free space on the eco-evolutionary dynamics}: For a comparative study, we plot the evolutions of $p$, $q$ and $r$ for (a) $\psi_1=\psi_2=\psi_3=0$, (b) $\psi_1=\psi_2=\psi_3=2$, (c) $\psi_1=2$, $\psi_2=\psi_3=0$, (d) $\psi_1=\psi_2=2$, $\psi_3=0$, (e) $\psi_1=\psi_3=0$, $\psi_2=2$, (f) $\psi_1=\psi_2=0$, $\psi_3=2$, (g) $\psi_1=\psi_3=2$, $\psi_2=0$, and (h) $\psi_2=\psi_3=2$, $\psi_1=0$. This figure demonstrates that free space strongly influences the emergent asymptotic dynamics of the model \eqref{16}. (a) In the absence of free space's contribution, prey can only survive. (b) While if free space contributes equally to everyone's fitness, then all can coexist. (c) If only free space benefits the prey, then prey and predator can coexist simultaneously. (d) All three can survive simultaneously if free space only contributes to prey and predator. (e) If free space facilitates the predator only, predator and parasite coexist. (f) The sole contribution of free space toward the parasite does not significantly differ from the null contribution of free space, at least for the chosen parameter set. (g) The equal contribution towards only prey and parasite allows the prey and predator to survive; however, the parasite dies in the long run. (h) If free space contributes only to predators and parasites, they all can coexist. Parameters: $\alpha=1.0$, $\beta=0.8$, $\epsilon=0.3$, $\gamma=0.4$, $\delta=1.0$, $\xi=0.5$, and $\eta=0.6$. Initial densities: $p_0=0.3$, $q_0=0.3$, and $r_0=0.3$.}
	\label{Fig7}
\end{figure*}
\subsection{Effect of prey's death rate} Also, the prey's death rate is a function of $\alpha$; thus, if $\alpha \in (0,1]$ increases, the death rate decreases for the prey. For $\alpha>1$, the death rate $(1-\alpha)$ for the prey is negative; hence it is not biologically meaningful. We plot the variation of dynamics in Fig.\ \eqref{Fig6} by varying $\alpha \in  (0,1.2]$ with a fixed step-length  $0.001$. For $\alpha>1.004$, the overall population $p+q+r$ exceeds the unity (See Fig.\ \eqref{Fig6} (d)). Within $\alpha \in  (0,1]$, the overall population $p+q+r$ remains always within the bounded domain $[0,1]$. At $\alpha=0.98402536$, the system \eqref{16} goes through a Hopf bifurcation, and the periodic solution arises. We notice prey's population remains extinct till $\alpha=0.58348744$. Beyond this value of $\alpha$, the prey's normalized density increases in the stationary state regime as anticipated (See Fig.\ \eqref{Fig6} (a)). The higher values of $\alpha \in (0,1]$ allow the prey to survive under favorable circumstances as the corresponding death rate decreases. The predator's density will also increase whenever the prey's density gets the opportunity to have enhancement, as reflected through Fig.\  \eqref{Fig6} (b). Interestingly, the parasite's density will diminish in the steady state regime for $\alpha \in (0.58348744,0.98402536)$ (See Fig.\ \eqref{Fig6} (c)). As the prey's density increases, they eat more and more insect parasites, and simultaneously this will reduce the parasite's density. 
\subsection{Importance of altruistic free space} To understand the role of philanthropic free space, we plot a few temporal evolutions of $p$, $q$, and $r$ in Fig.\ \eqref{Fig7} for different values of $\psi_1$, $\psi_2$, and $\psi_3$.  We also set the initial condition at $(0.3,0.3,0.3)$ for all these eight subfigures. Figure \eqref{Fig7} (a) depicts that only prey can survive alone if free space does not contribute altruistically ($\psi_1=\psi_2=\psi_3=0$) to anybody for the particular chosen parameter set. To get a comparative understanding, we set $\psi_1=\psi_2=\psi_3=2$ in Fig.\ \eqref{Fig7} (b). Thus, all species will get an equal amount of benefits from the free space. Clearly, Fig.\ \eqref{Fig7} (b) portrays the coexistence of all three, and we find $q>r>p$ inequality in the asymptotic limit. Thus, predators dominate the other two species for this chosen parameter set. In Fig.\ \eqref{Fig7} (c), we choose $\psi_1=2$ and $\psi_2=\psi_3=0$. This indicates that the free space will only entertain the prey. Interestingly, while in the absence of any positive contribution from free space, Fig.\ \eqref{Fig7} (a) depicts the sole survivability of the prey; here, in Fig.\ \eqref{Fig7}(c), we find the coexistence of both prey and predator. This attests to the influential contribution of free space in our constructed model. All three species can co-exist if the free space further provides a positive, generous contribution to the predator. Figure\ \eqref{Fig7} (d) is drawn with $\psi_1=\psi_2=2$ and $\psi_3=0$. This simultaneous appearance of all three prey, predator, and parasite is not observed in the system \eqref{3}. Now, if the free space will entertain only the predator, one may observe a different stationary point in the long run. We choose $\psi_1=\psi_3=0$ and $\psi_2=2$ in Fig.\ \eqref{Fig7} (e). This allows the system to converge in the prey-free stationary state. However, when all $\psi_1=\psi_2=\psi_3=0$, prey are the sole survivor as per Fig.\ \eqref{Fig7} (a). As soon as free space provides a selfless contribution to only the predators, the prey vanishes from society. Nevertheless, if free space favors the parasites alone, that does not make any significant difference in nature. We choose $\psi_1=\psi_2=0$ and $\psi_3=2$ in Fig.\ \eqref{Fig7} (f). This will again lead to a prey dominated society free from predators as well as parasites. Prey can also dominate the society if free space favors both prey and parasites. Figure\ \eqref{Fig7} (g) reflects the concurrence of prey and predator in a parasite-free society. This subfigure is drawn for $\psi_1=\psi_3=2$ and $\psi_2=0$. Prey dominate the predator in the society for this parameter set. Interestingly, prey is dominated by other two if free space will not contribute in the prey's fitness. For $\psi_1=0$ and $\psi_2=\psi_3=2$, all predator-prey-parasites co-exist as observed in Fig.\ \eqref{Fig7} (h). Despite the free space acts like a selfless entity in our model, its contribution to the co-evolution of all species is massive, as illustrated through this figure.

\section{Conclusions} \label{Conclusion}

 There exists a vast literature dealing with predator-prey interaction. All these model variants try to capture a thorough understanding of the underlying microscopic processes of ecological species interaction. Our study on predator-prey-parasite interaction explores many insightful results on biological systems. The ecological signature of free space in our theoretical model allows the coexistence of all three species and, thus, plays an assertive role in the maintenance of biodiversity in nature. The consideration of free space's charitable role promotes biodiversity sustenance, which is impossible with a model formed analogously to the Lotka-Volterra model. Our numerical simulations, along with the analytical findings, support this understanding. We derive both systems' stationary points' existence, uniqueness, positivity, and local stability criteria analytically. We are further able to capture the beauty of cyclic dominance among prey, predator, and parasite. This cyclic dominance supports the realistic understanding that there is no sure-shot winner in the long run. Instead, one may dominate the others in a specific time window; however, it is dominated by others in a different time window. This cyclic dominance lies at the heart of species coexistence and the maintenance of biodiversity.
 
\par {Note that, despite such substantial existing literature on this topic, the selfless contribution of free space in nurturing the evolutionary scenarios is neglected in most studies to the best of our knowledge. We thus hope our simple three-dimensional eco-evolutionary model may serve as a fundamental stepping stone toward this research direction. We further emphasize that our proposed nonlinear system is far from practical physical scenarios, as our mathematical model consists of some simplifying assumptions. Still, such a simplified model can illuminate novel dynamical phenomena depicting several valuable information.  Our mathematical model may provide further exciting outcomes if one adds additional components like environmental fluctuations} \cite{may1973stability}{, delay} \cite{schwartz2015noise}{, the presence of fear factors} \cite{creel2008relationships,biswas2021delay}{, different food sources for the predators} \cite{ghosh2017prey}{, and many more. In fact, we consider only the pairwise interaction in the payoff matrix. One may consider higher-order interaction} \cite{chatterjee2022controlling,majhi2022dynamics} {and can anticipate more diverse emerging dynamical states. It isn't accessible to claim any biological applications immediately of the model studied here. Nevertheless, the cyclic dominance among competing species in the form of periodic dynamics motivates us to report the current eco-evolutionary model proposed on theoretical grounds.} 
\par {Despite various limitations, our proposed theoretical model can be generalized to several patches, and such different network topologies} \cite{newman2018networks} {may contain diverse possible time-invariant and time-varying connectivities} \cite{li2020evolution,chowdhury2019convergence,ghosh2022synchronized,chowdhury2019synchronization1,holme2012temporal,dixit2021emergent,dixit2021dynamic}. {Studying the role of a network structure using dynamical systems on collective behavior} \cite{jusup2022social,chowdhury2022extreme,parastesh2021chimeras,chowdhury2020effect,wang2016statistical,perc2017statistical,khaleghi2019chimera,kabir2020impact} {gains wide recognition due to its two-fold appeal. Firstly, it will allow grasping a better understanding of several natural phenomena. Secondly, it provides a handy entry point for devising efficient performing devices from the technological point of view. On the other hand, organisms' active and passive dispersal can substantially affect ecological dynamics, as reported earlier in various Refs.} \cite{peltomaki2008three,comins1996persistence,dieckmann1999evolutionary,ellner2001habitat,baguette2012dispersal,crowley1981dispersal,hassell1974aggregation}. {The investigation by Holland et al.} \cite{holland2008strong} {demonstrates that irregularities in connectivities among different patches (sites) of a dispersal network of predators and prey generally offer prolonged transient and are favorable for asynchronous ecological dynamics, leading to lower amplitude fluctuations in population abundances. A recent study}  \cite{su2022evolution} {suggests altruistic unidirectional behavior of individuals can facilitate and promote cooperation in social networks. Thus, the earlier investigations suggest that the consequence of various network structures and dispersal dynamics will eventually lead to more exciting dynamics, and these explorations remain an interesting core avenue for future research. Observing how our theoretical framework allows a possible new range of insights while generalized to classical three- or four-strategy cyclic games} \cite{intoy2015synchronization,szolnoki2020pattern,kabir2021cyclic,mobilia2010oscillatory,islam2022effect,bazeia2018phase,szolnoki2015vortices,mathiesen2011ecosystems,berr2009zero} {will be further interesting.} We hope that our theoretical investigation of the mathematical model may advance our understanding of social diversity and probably inspire as well as motivate at least a few readers to shed light on the predator-prey interplay of ecology and evolutionary dynamics.

\section*{Data accessibility.} 
This article has no additional data.
\section*{Authors’ contributions (CRediT authorship contribution statement).} 
{\bf Sayantan Nag Chowdhury:} Conceptualization, Methodology, Software, Validation, Formal analysis, Investigation, Resources, Data Curation, Writing - Original Draft, Visualization, Supervision, Project administration. {\bf Jeet Banerjee:} Conceptualization, Methodology, Validation, Formal analysis, Writing - Review \& Editing. {\bf Matja{\v z} Perc:} Validation, Resources, Writing - Review \& Editing, Visualization, Supervision, Project administration, Funding acquisition. {\bf Dibakar Ghosh:} Validation, Data Curation, Writing - Review \& Editing, Visualization, Supervision, Project administration.  
\section*{Competing interests.} We declare we have no competing interests.
\section*{Funding.}  M.P. was supported by the Slovenian Research Agency (grant nos. P1-0403 and J1-2457).
\section*{Supplementary material}
The supplementary material associated with this article contains the stationary points of the systems \eqref{3} and \eqref{16}.

\bibliographystyle{elsarticle-num}  
\bibliography{Jeet}

\providecommand{\noopsort}[1]{}\providecommand{\singleletter}[1]{#1}%
\begin{thebibliography}{10}
\expandafter\ifx\csname url\endcsname\relax
  \def\url#1{\texttt{#1}}\fi
\expandafter\ifx\csname urlprefix\endcsname\relax\def\urlprefix{URL }\fi
\expandafter\ifx\csname href\endcsname\relax
  \def\href#1#2{#2} \def\path#1{#1}\fi

\bibitem{malthus2007essay}
T.~R. Malthus, An essay on the principle of population, as it affects the
  future imporvement of society, with remarks on the speculations of Mr.
  Godwin, M. Condorcet, and other writers, The Lawbook Exchange, Ltd., 2007.

\bibitem{lotka1925elements}
A.~J. Lotka, Elements of physical biology, Williams \& Wilkins, 1925.

\bibitem{volterra1926variazioni}
V.~Volterra, Variazioni e fluttuazioni del numero d'individui in specie animali
  conviventi, Societ{\`a} anonima tipografica" Leonardo da Vinci", 1926.

\bibitem{rosenzweig1963graphical}
M.~L. Rosenzweig, R.~H. MacArthur, Graphical representation and stability
  conditions of predator-prey interactions, The American Naturalist 97~(895)
  (1963) 209--223.

\bibitem{chowdhury2021complex}
S.~Nag~Chowdhury, S.~Kundu, M.~Perc, D.~Ghosh, Complex evolutionary dynamics
  due to punishment and free space in ecological multigames, Proceedings of the
  Royal Society A 477~(2252) (2021) 20210397.

\bibitem{chowdhury2021eco}
S.~Nag~Chowdhury, S.~Kundu, J.~Banerjee, M.~Perc, D.~Ghosh, Eco-evolutionary
  dynamics of cooperation in the presence of policing, Journal of Theoretical
  Biology 518 (2021) 110606.

\bibitem{roy2022eco}
S.~Roy, S.~Nag~Chowdhury, P.~C. Mali, M.~Perc, D.~Ghosh, Eco-evolutionary
  dynamics of multigames with mutations, Plos one 17~(8) (2022) e0272719.

\bibitem{helbing2009outbreak}
D.~Helbing, W.~Yu, The outbreak of cooperation among success-driven individuals
  under noisy conditions, Proceedings of the National Academy of Sciences
  106~(10) (2009) 3680--3685.

\bibitem{nag2020cooperation}
S.~Nag~Chowdhury, S.~Kundu, M.~Duh, M.~Perc, D.~Ghosh, Cooperation on
  interdependent networks by means of migration and stochastic imitation,
  Entropy 22~(4) (2020) 485.

\bibitem{jiang2010role}
L.-L. Jiang, W.-X. Wang, Y.-C. Lai, B.-H. Wang, Role of adaptive migration in
  promoting cooperation in spatial games, Physical Review E 81~(3) (2010)
  036108.

\bibitem{chowdhury2020distance}
S.~Nag~Chowdhury, S.~Majhi, D.~Ghosh, Distance dependent competitive
  interactions in a frustrated network of mobile agents, IEEE Transactions on
  Network Science and Engineering 7~(4) (2020) 3159--3170.

\bibitem{meloni2009effects}
S.~Meloni, A.~Buscarino, L.~Fortuna, M.~Frasca, J.~G{\'o}mez-Garde{\~n}es,
  V.~Latora, Y.~Moreno, Effects of mobility in a population of prisoner’s
  dilemma players, Physical Review E 79~(6) (2009) 067101.

\bibitem{Sar_2022}
G.~K. Sar, S.~{Nag Chowdhury}, M.~Perc, D.~Ghosh, Swarmalators under
  competitive time-varying phase interactions, New Journal of Physics 24~(4)
  (2022) 043004.

\bibitem{noh2004random}
J.~D. Noh, H.~Rieger, Random walks on complex networks, Physical Review Letters
  92~(11) (2004) 118701.

\bibitem{aktipis2004know}
C.~A. Aktipis, Know when to walk away: contingent movement and the evolution of
  cooperation, Journal of Theoretical Biology 231~(2) (2004) 249--260.

\bibitem{vainstein2007does}
M.~H. Vainstein, A.~T. Silva, J.~J. Arenzon, Does mobility decrease
  cooperation?, Journal of Theoretical Biology 244~(4) (2007) 722--728.

\bibitem{chowdhury2019synchronization}
S.~Nag~Chowdhury, S.~Majhi, M.~Ozer, D.~Ghosh, M.~Perc, Synchronization to
  extreme events in moving agents, New Journal of Physics 21~(7) (2019) 073048.

\bibitem{smaldino2012movement}
P.~E. Smaldino, J.~C. Schank, Movement patterns, social dynamics, and the
  evolution of cooperation, Theoretical Population Biology 82~(1) (2012)
  48--58.

\bibitem{szolnoki2014cyclic}
A.~Szolnoki, M.~Mobilia, L.-L. Jiang, B.~Szczesny, A.~M. Rucklidge, M.~Perc,
  Cyclic dominance in evolutionary games: a review, Journal of the Royal
  Society Interface 11~(100) (2014) 20140735.

\bibitem{nahum2011evolution}
J.~R. Nahum, B.~N. Harding, B.~Kerr, Evolution of restraint in a structured
  rock--paper--scissors community, Proceedings of the National Academy of
  Sciences 108~(supplement\_2) (2011) 10831--10838.

\bibitem{kerr2002local}
B.~Kerr, M.~A. Riley, M.~W. Feldman, B.~J. Bohannan, Local dispersal promotes
  biodiversity in a real-life game of rock--paper--scissors, Nature 418~(6894)
  (2002) 171--174.

\bibitem{lankau2007mutual}
R.~A. Lankau, S.~Y. Strauss, Mutual feedbacks maintain both genetic and species
  diversity in a plant community, Science 317~(5844) (2007) 1561--1563.

\bibitem{durrett1998spatial}
R.~Durrett, S.~Levin, Spatial aspects of interspecific competition, Theoretical
  Population Biology 53~(1) (1998) 30--43.

\bibitem{jackson1975alleopathy}
J.~Jackson, L.~Buss, Alleopathy and spatial competition among coral reef
  invertebrates, Proceedings of the National Academy of Sciences 72~(12) (1975)
  5160--5163.

\bibitem{elowitz2000synthetic}
M.~B. Elowitz, S.~Leibler, A synthetic oscillatory network of transcriptional
  regulators, Nature 403~(6767) (2000) 335--338.

\bibitem{sinervo1996rock}
B.~Sinervo, C.~M. Lively, The rock--paper--scissors game and the evolution of
  alternative male strategies, Nature 380~(6571) (1996) 240--243.

\bibitem{gilg2003cyclic}
O.~Gilg, I.~Hanski, B.~Sittler, Cyclic dynamics in a simple vertebrate
  predator-prey community, Science 302~(5646) (2003) 866--868.

\bibitem{guill2011three}
C.~Guill, B.~Drossel, W.~Just, E.~Carmack, A three-species model explaining
  cyclic dominance of pacific salmon, Journal of Theoretical Biology 276~(1)
  (2011) 16--21.

\bibitem{hofbauer1998evolutionary}
J.~Hofbauer, K.~Sigmund, et~al., Evolutionary games and population dynamics,
  Cambridge university press, 1998.

\bibitem{nowak2006evolutionary}
M.~A. Nowak, Evolutionary dynamics: exploring the equations of life, Harvard
  university press, 2006.

\bibitem{reichenbach2007noise}
T.~Reichenbach, M.~Mobilia, E.~Frey, Noise and correlations in a spatial
  population model with cyclic competition, Physical Review Letters 99~(23)
  (2007) 238105.

\bibitem{he2010spatial}
Q.~He, M.~Mobilia, U.~C. T{\"a}uber, Spatial rock-paper-scissors models with
  inhomogeneous reaction rates, Physical Review E 82~(5) (2010) 051909.

\bibitem{kabir2021role}
K.~A. Kabir, J.~Tanimoto, The role of pairwise nonlinear evolutionary dynamics
  in the rock--paper--scissors game with noise, Applied Mathematics and
  Computation 394 (2021) 125767.

\bibitem{reichenbach2007mobility}
T.~Reichenbach, M.~Mobilia, E.~Frey, Mobility promotes and jeopardizes
  biodiversity in rock--paper--scissors games, Nature 448~(7157) (2007)
  1046--1049.

\bibitem{hauert2006evolutionary}
C.~Hauert, M.~Holmes, M.~Doebeli, Evolutionary games and population dynamics:
  maintenance of cooperation in public goods games, Proceedings of the Royal
  Society B: Biological Sciences 273~(1600) (2006) 2565--2571.

\bibitem{gokhale2016eco}
C.~S. Gokhale, C.~Hauert, Eco-evolutionary dynamics of social dilemmas,
  Theoretical Population Biology 111 (2016) 28--42.

\bibitem{dhooge2003matcont}
A.~Dhooge, W.~Govaerts, Y.~A. Kuznetsov, Matcont: a matlab package for
  numerical bifurcation analysis of odes, ACM Transactions on Mathematical
  Software (TOMS) 29~(2) (2003) 141--164.

\bibitem{hale2012dynamics}
J.~K. Hale, H.~Ko{\c{c}}ak, Dynamics and bifurcations, Vol.~3, Springer Science
  \& Business Media, 2012.

\bibitem{may1973stability}
R.~M. May, Stability in randomly fluctuating versus deterministic environments,
  The American Naturalist 107~(957) (1973) 621--650.

\bibitem{schwartz2015noise}
I.~B. Schwartz, L.~Billings, T.~W. Carr, M.~Dykman, Noise-induced switching and
  extinction in systems with delay, Physical Review E 91~(1) (2015) 012139.

\bibitem{creel2008relationships}
S.~Creel, D.~Christianson, Relationships between direct predation and risk
  effects, Trends in Ecology \& Evolution 23~(4) (2008) 194--201.

\bibitem{biswas2021delay}
S.~Biswas, P.~K. Tiwari, S.~Pal, Delay-induced chaos and its possible control
  in a seasonally forced eco-epidemiological model with fear effect and
  predator switching, Nonlinear Dynamics 104~(3) (2021) 2901--2930.

\bibitem{ghosh2017prey}
J.~Ghosh, B.~Sahoo, S.~Poria, Prey-predator dynamics with prey refuge providing
  additional food to predator, Chaos, Solitons \& Fractals 96 (2017) 110--119.

\bibitem{chatterjee2022controlling}
S.~Chatterjee, S.~Nag~Chowdhury, D.~Ghosh, C.~Hens, Controlling species
  densities in structurally perturbed intransitive cycles with higher-order
  interactions, Chaos 32 (2022) 103122.

\bibitem{majhi2022dynamics}
S.~Majhi, M.~Perc, D.~Ghosh, Dynamics on higher-order networks: A review,
  Journal of the Royal Society Interface 19~(188) (2022) 20220043.

\bibitem{newman2018networks}
M.~Newman, Networks, Oxford university press, 2018.

\bibitem{li2020evolution}
A.~Li, L.~Zhou, Q.~Su, S.~P. Cornelius, Y.-Y. Liu, L.~Wang, S.~A. Levin,
  Evolution of cooperation on temporal networks, Nature Communications 11~(1)
  (2020) 2259.

\bibitem{chowdhury2019convergence}
S.~Nag~Chowdhury, S.~Majhi, D.~Ghosh, A.~Prasad, Convergence of chaotic
  attractors due to interaction based on closeness, Physics Letters A 383~(35)
  (2019) 125997.

\bibitem{ghosh2022synchronized}
D.~Ghosh, M.~Frasca, A.~Rizzo, S.~Majhi, S.~Rakshit, K.~Alfaro-Bittner,
  S.~Boccaletti, The synchronized dynamics of time-varying networks, Physics
  Reports 949 (2022) 1--63.

\bibitem{chowdhury2019synchronization1}
S.~Nag~Chowdhury, D.~Ghosh, Synchronization in dynamic network using threshold
  control approach, Europhysics Letters 125~(1) (2019) 10011.

\bibitem{holme2012temporal}
P.~Holme, J.~Saram{\"a}ki, Temporal networks, Physics Reports 519~(3) (2012)
  97--125.

\bibitem{dixit2021emergent}
S.~Dixit, S.~Nag~Chowdhury, A.~Prasad, D.~Ghosh, M.~D. Shrimali, Emergent
  rhythms in coupled nonlinear oscillators due to dynamic interactions, Chaos:
  An Interdisciplinary Journal of Nonlinear Science 31~(1) (2021) 011105.

\bibitem{dixit2021dynamic}
S.~Dixit, S.~Nag~Chowdhury, D.~Ghosh, M.~D. Shrimali, Dynamic interaction
  induced explosive death, Europhysics Letters 133~(4) (2021) 40003.

\bibitem{jusup2022social}
M.~Jusup, P.~Holme, K.~Kanazawa, M.~Takayasu, I.~Romi{\'c}, Z.~Wang,
  S.~Ge{\v{c}}ek, T.~Lipi{\'c}, B.~Podobnik, L.~Wang, et~al., Social physics,
  Physics Reports 948 (2022) 1--148.

\bibitem{chowdhury2022extreme}
S.~Nag~Chowdhury, A.~Ray, S.~K. Dana, D.~Ghosh, Extreme events in dynamical
  systems and random walkers: A review, Physics Reports 966 (2022) 1--52.

\bibitem{parastesh2021chimeras}
F.~Parastesh, S.~Jafari, H.~Azarnoush, Z.~Shahriari, Z.~Wang, S.~Boccaletti,
  M.~Perc, Chimeras, Physics Reports 898 (2021) 1--114.

\bibitem{chowdhury2020effect}
S.~Nag~Chowdhury, D.~Ghosh, C.~Hens, Effect of repulsive links on frustration
  in attractively coupled networks, Physical Review E 101~(2) (2020) 022310.

\bibitem{wang2016statistical}
Z.~Wang, C.~T. Bauch, S.~Bhattacharyya, A.~d'Onofrio, P.~Manfredi, M.~Perc,
  N.~Perra, M.~Salath{\'e}, D.~Zhao, Statistical physics of vaccination,
  Physics Reports 664 (2016) 1--113.

\bibitem{perc2017statistical}
M.~Perc, J.~J. Jordan, D.~G. Rand, Z.~Wang, S.~Boccaletti, A.~Szolnoki,
  Statistical physics of human cooperation, Physics Reports 687 (2017) 1--51.

\bibitem{khaleghi2019chimera}
L.~Khaleghi, S.~Panahi, S.~Nag~Chowdhury, S.~Bogomolov, D.~Ghosh, S.~Jafari,
  Chimera states in a ring of map-based neurons, Physica A: Statistical
  Mechanics and its Applications 536 (2019) 122596.

\bibitem{kabir2020impact}
K.~A. Kabir, K.~Kuga, J.~Tanimoto, The impact of information spreading on
  epidemic vaccination game dynamics in a heterogeneous complex network-a
  theoretical approach, Chaos, Solitons \& Fractals 132 (2020) 109548.

\bibitem{peltomaki2008three}
M.~Peltom{\"a}ki, M.~Alava, Three-and four-state rock-paper-scissors games with
  diffusion, Physical Review E 78~(3) (2008) 031906.

\bibitem{comins1996persistence}
H.~a. Comins, M.~Hassell, Persistence of multispecies host--parasitoid
  interactions in spatially distributed models with local dispersal, Journal of
  Theoretical Biology 183~(1) (1996) 19--28.

\bibitem{dieckmann1999evolutionary}
U.~Dieckmann, B.~O'Hara, W.~Weisser, The evolutionary ecology of dispersal,
  Trends in Ecology \& Evolution 14~(3) (1999) 88--90.

\bibitem{ellner2001habitat}
S.~P. Ellner, E.~McCauley, B.~E. Kendall, C.~J. Briggs, P.~R. Hosseini, S.~N.
  Wood, A.~Janssen, M.~W. Sabelis, P.~Turchin, R.~M. Nisbet, et~al., Habitat
  structure and population persistence in an experimental community, Nature
  412~(6846) (2001) 538--543.

\bibitem{baguette2012dispersal}
M.~Baguette, T.~G. Benton, J.~M. Bullock, Dispersal ecology and evolution,
  Oxford University Press, 2012.

\bibitem{crowley1981dispersal}
P.~H. Crowley, Dispersal and the stability of predator-prey interactions, The
  American Naturalist 118~(5) (1981) 673--701.

\bibitem{hassell1974aggregation}
M.~P. Hassell, R.~May, Aggregation of predators and insect parasites and its
  effect on stability, The Journal of Animal Ecology (1974) 567--594.

\bibitem{holland2008strong}
M.~D. Holland, A.~Hastings, Strong effect of dispersal network structure on
  ecological dynamics, Nature 456~(7223) (2008) 792--794.

\bibitem{su2022evolution}
Q.~Su, B.~Allen, J.~B. Plotkin, Evolution of cooperation with asymmetric social
  interactions, Proceedings of the National Academy of Sciences 119~(1) (2022)
  e2113468118.

\bibitem{intoy2015synchronization}
B.~Intoy, M.~Pleimling, Synchronization and extinction in cyclic games with
  mixed strategies, Physical Review E 91~(5) (2015) 052135.

\bibitem{szolnoki2020pattern}
A.~Szolnoki, B.~De~Oliveira, D.~Bazeia, Pattern formations driven by cyclic
  interactions: A brief review of recent developments, Europhysics Letters
  131~(6) (2020) 68001.

\bibitem{kabir2021cyclic}
K.~A. Kabir, J.~Tanimoto, A cyclic epidemic vaccination model: Embedding the
  attitude of individuals toward vaccination into svis dynamics through social
  interactions, Physica A: Statistical Mechanics and its Applications 581
  (2021) 126230.

\bibitem{mobilia2010oscillatory}
M.~Mobilia, Oscillatory dynamics in rock--paper--scissors games with mutations,
  Journal of Theoretical Biology 264~(1) (2010) 1--10.

\bibitem{islam2022effect}
S.~Islam, A.~Mondal, M.~Mobilia, S.~Bhattacharyya, C.~Hens, Effect of mobility
  in the rock-paper-scissor dynamics with high mortality, Physical Review E
  105~(1) (2022) 014215.

\bibitem{bazeia2018phase}
D.~Bazeia, B.~de~Oliveira, A.~Szolnoki, Phase transitions in dependence of apex
  predator decaying ratio in a cyclic dominant system, Europhysics Letters
  124~(6) (2018) 68001.

\bibitem{szolnoki2015vortices}
A.~Szolnoki, M.~Perc, Vortices determine the dynamics of biodiversity in
  cyclical interactions with protection spillovers, New Journal of Physics
  17~(11) (2015) 113033.

\bibitem{mathiesen2011ecosystems}
J.~Mathiesen, N.~Mitarai, K.~Sneppen, A.~Trusina, Ecosystems with mutually
  exclusive interactions self-organize to a state of high diversity, Physical
  Review Letters 107~(18) (2011) 188101.

\bibitem{berr2009zero}
M.~Berr, T.~Reichenbach, M.~Schottenloher, E.~Frey, Zero-one survival behavior
  of cyclically competing species, Physical Review Letters 102~(4) (2009)
  048102.

\end{thebibliography}

\end{document}